\documentclass[zpreprint,zbstpl]{./LaTeX/zeus/zeus_paper}
\usepackage[english]{babel}
\newcommand{\Zdetdesc}{%
A detailed description of the ZEUS detector can be found 
elsewhere~\cite{zeus:1993:bluebook}. A brief outline of the 
components that are most relevant for this analysis is given
below.\xspace}

\chardef\usc=95
\chardef\til=126
\catcode`\@=11 
\DeclareRobustCommand\xdotspace{\futurelet\@let@token\@xdotspace}
\def\@xdotspace{%
  \ifx\@let@token.\else
  \ifx\@let@token\bgroup.\else
  \ifx\@let@token\egroup.\else
  \ifx\@let@token\/.\else
  \ifx\@let@token\ .\else
  \ifx\@let@token~.\else
  \ifx\@let@token!.\else
  \ifx\@let@token,.\else
  \ifx\@let@token:.\else
  \ifx\@let@token;.\else
  \ifx\@let@token?.\else
  \ifx\@let@token/.\else
  \ifx\@let@token'.\else
  \ifx\@let@token).\else
  \ifx\@let@token-.\else
  \ifx\@let@token\@xobeysp.\else
  \ifx\@let@token\space.\else
  \ifx\@let@token\@sptoken.\else
   .\space
   \fi\fi\fi\fi\fi\fi\fi\fi\fi\fi\fi\fi\fi\fi\fi\fi\fi\fi}
\catcode`\@=12 

\newcommand{\stru}[2]{%
   \relax\ifmmode\hbox{\vrule height#1 depth#2 width0pt}%
   \else\vrule height#1 depth#2 width0pt\fi}

\newcommand{\Ronum}[1]{\uppercase\expandafter{\romannumeral#1}}
\newcommand{\ronum}[1]{\expandafter{\romannumeral#1}}
\DeclareRobustCommand{\LaTeXZ}{%
  \LaTeX\kern-.05em4\kern-.1em
  {\raisebox{-0.2ex}{$\scriptstyle\text{ZEUS}$}}\xspace}

\newcommand{\tab}[1]{Table~\ref{tab-#1}}

\DeclareMathAlphabet{\mathbf}{OT1}{cmr}{bx}{sl}
\newcommand{\eVdist}{\kern-0.06667em}

\newcommand{\gev}{{\,\text{Ge}\eVdist\text{V\/}}}

\newcommand{\fb}{\,\text{fb}}

\newcommand{\pbi}{\,\text{pb}^{-1}}

\newcommand{\cm}{\,\text{cm}}

\newcommand{\Tesla}{\,\text{T}}

\newcommand{\slashfrac}[2]{%
  \raisebox{0.5ex}{\ensuremath #1}\kern-0.12em/\kern-0.08em
  \raisebox{-.8ex}{\ensuremath #2}}

\newcommand{\sqr}[3]{%
    {\vcenter{\hrule height.#3ex\hbox{\vrule width.#2ex height#1ex
     \kern#1ex\vrule width.#3ex}\hrule height.#2ex}}}

\catcode`\@=11 
\newcommand{\parenbar}{\mathpalette\p@renb@r}
\def\p@renb@r#1#2{\vbox{%
  \ifx#1\scriptscriptstyle \dimen@.7em\dimen@ii.2em\else
  \ifx#1\scriptstyle \dimen@.8em\dimen@ii.25em\else
  \dimen@1em\dimen@ii.4em\fi\fi \offinterlineskip
  \ialign{\hfill##\hfill\cr
    \vbox{\hrule width\dimen@ii}\cr
    \noalign{\vskip-.3ex}%
    \hbox to\dimen@{$\mathchar300\hfil\mathchar301$}\cr
    \noalign{\vskip-.3ex}%
    $#1#2$\cr}}}
\catcode`\@=12 

\newcommand{\IP}{{\rm I$\kern-0.01667em$P}\xspace}

\mathchardef\qsm=63
\mathchardef\pls=43
\mathchardef\mns=512
\mathchardef\plm=518
\mathchardef\eql=61
\mathchardef\smallleft=300
\mathchardef\smallright=301
\mathchardef\les=316
\mathchardef\gre=318
\mathchardef\leq=532
\mathchardef\grq=533

\catcode`\@=11 
\newcounter{pict@width}
\newcounter{pict@height}
\newlength{\pict@scale}
\newcommand{\psfigadd}[4]{%
\setcounter{pict@width}{1*\ratio{#2+\pict@scale/2}{\pict@scale}}
\setcounter{pict@height}{1*\ratio{#3+\pict@scale/2}{\pict@scale}}
\setlength{\unitlength}{\pict@scale}
\hbox to #2{\hspace{-\fill}\begin{picture}(\thepict@width,\thepict@height)
\put(0,0){\psfig{figure=#1,width=#2,height=#3,clip=}}
\SetScale{0.283466457}
\SetWidth{1.763889}
{#4}
\end{picture}}
}
\newcounter{pict@widthfst}
\newcounter{pict@widthscd}
\newcounter{pict@widthtot}
\newcommand{\psfigaddtwo}[7]{%
\setcounter{pict@widthfst}{1*\ratio{#2+\pict@scale/2}{\pict@scale}}
\setcounter{pict@widthscd}{1*\ratio{#2+#4+\pict@scale/2}{\pict@scale}}
\setcounter{pict@widthtot}{1*\ratio{#2+#4+#6+\pict@scale/2}{\pict@scale}}
\setcounter{pict@height}{1*\ratio{#3+\pict@scale/2}{\pict@scale}}
\setlength{\unitlength}{\pict@scale}
\hbox{\hspace{-\fill}\begin{picture}(\thepict@widthtot,\thepict@height)
\put(0,0){\psfig{figure=#1,width=#2,height=#3,clip=}}
\put(\thepict@widthscd,0){\psfig{figure=#5,width=#6,height=#3,clip=}}
\SetScale{0.283466457}
\SetWidth{1.763889}
{#7}
\end{picture}}
}
\newcommand{\psfigror}[4]{%
\setcounter{pict@width}{1*\ratio{#2+\pict@scale/2}{\pict@scale}}
\setcounter{pict@height}{1*\ratio{#3+\pict@scale/2}{\pict@scale}}
\setlength{\unitlength}{\pict@scale}
\hbox{\begin{picture}(\thepict@width,\thepict@height)
\put(0,\thepict@height){\psfig{figure=#1,width=#3,height=#2,clip=,angle=270}}
\SetScale{0.283466457}
\SetWidth{1.763889}
{#4}
\end{picture}}
}
\newcommand{\psfigrol}[4]{%
\setcounter{pict@width}{1*\ratio{#2+\pict@scale/2}{\pict@scale}}
\setcounter{pict@height}{1*\ratio{#3+\pict@scale/2}{\pict@scale}}
\setlength{\unitlength}{\pict@scale}
\hbox{\begin{picture}(\thepict@width,\thepict@height)
\put(0,0){\psfig{figure=#1,width=#3,height=#2,clip=,angle=90}}
\SetScale{0.283466457}
\SetWidth{1.763889}
{#4}
\end{picture}}
}
\catcode`\@=12 
\newlength\listtextwidth

\catcode`\@=11 
\newlength{\@tabfninsert}
\newlength{\@tabfnwidth}
\newcommand{\tabfootnote}[2]{%
  \setlength{\@tabfninsert}{0.8em}
  \setlength{\@tabfnwidth}{\textwidth}
  \addtolength{\@tabfnwidth}{-\@tabfninsert}
  \addtolength{\@tabfnwidth}{-0.4em}
  \noindent\makebox[\@tabfninsert][r]{\footnotesize$^{#1}$\hfil}\hfill%
  \parbox[t]{\@tabfnwidth}{\footnotesize #2\hfill}}
\catcode`\@=12 

\definecolor{darkyellow}{rgb}{0.95,0.95,0.0}

\newcommand{\bec}{\begin{center}}
\newcommand{\bef}{\begin{figure}}
\newcommand{\bet}{\begin{table}}
\newcommand{\bee}{\begin{equation}}
\newcommand{\bei}{\begin{itemize}}
\newcommand{\enc}{\end{center}}
\newcommand{\enf}{\end{figure}}
\newcommand{\ent}{\end{table}}
\newcommand{\ene}{\end{equation}}
\newcommand{\eni}{\end{itemize}}

\def\citeCTD{{\cite{%
nim:a279:290,*npps:b32:181,*nim:a338:254%
}}\xspace}
\def\citeCAL{{\cite{%
nim:a309:77,*nim:a309:101,*nim:a321:356,*nim:a336:23%
}}\xspace}

\begin{document}
\newcommand{\massmu}{M_{\mu \mu}}
\prepnum{DESY-09-072}

\title{
Multi-lepton production at high transverse momentum at HERA
}                                                       
                    
\author{ZEUS Collaboration}
\date{June 2009}

\abstract{
A search for events containing two or more 
high-transverse-momentum isolated leptons has 
 been performed in $ep$ collisions with the ZEUS detector at HERA
 using the full collected data sample, corresponding to an integrated luminosity of $480 \pbi$.
 The number of observed events 
 has been compared with the prediction from the Standard Model, 
 searching for possible deviations, especially for multi-lepton
 events with invariant mass larger than 100~\gev. Good agreement with the 
 Standard Model has been observed.
 Total and differential cross sections for di-lepton production have been 
 measured in a restricted phase space dominated by photon-photon collisions.
}

\makezeustitle

\begin{center}                                                                                     
{                      \Large  The ZEUS Collaboration              }                               
\end{center}                                                                                       
  S.~Chekanov,                                                                                     
  M.~Derrick,                                                                                      
  S.~Magill,                                                                                       
  B.~Musgrave,                                                                                     
  D.~Nicholass$^{   1}$,                                                                           
  \mbox{J.~Repond},                                                                                
  R.~Yoshida\\                                                                                     
 {\it Argonne National Laboratory, Argonne, Illinois 60439-4815, USA}~$^{n}$                       
\par \filbreak                                                                                     
  M.C.K.~Mattingly \\                                                                              
 {\it Andrews University, Berrien Springs, Michigan 49104-0380, USA}                               
\par \filbreak                                                                                     
  P.~Antonioli,                                                                                    
  G.~Bari,                                                                                         
  L.~Bellagamba,                                                                                   
  D.~Boscherini,                                                                                   
  A.~Bruni,                                                                                        
  G.~Bruni,                                                                                        
  F.~Cindolo,                                                                                      
  M.~Corradi,                                                                                      
\mbox{G.~Iacobucci},                                                                               
  A.~Margotti,                                                                                     
  R.~Nania,                                                                                        
  A.~Polini\\                                                                                      
  {\it INFN Bologna, Bologna, Italy}~$^{e}$
% ...
\newpage                                                                                           
\enlargethispage{5cm}                                                                              
$^{\    1}$ also affiliated with University College London,                                        
United Kingdom\\                                                                                   
% ...
$^{\    5}$ now at Institute of Aviation, Warsaw, Poland \\                                        
% ...
%
\newpage   
                                                           %
                                                           %
\begin{tabular}[h]{rp{14cm}}                                                                       
$^{a}$ &  supported by the Natural Sciences and Engineering Research Council of Canada (NSERC) \\  
$^{b}$ &  supported by the German Federal Ministry for Education and Research (BMBF), under        
          contract Nos. 05 HZ6PDA, 05 HZ6GUA, 05 HZ6VFA and 05 HZ4KHA\\                            
$^{c}$ &  supported in part by the MINERVA Gesellschaft f\"ur Forschung GmbH, the Israel Science   
          Foundation (grant No. 293/02-11.2) and the US-Israel Binational Science Foundation \\    
$^{d}$ &  supported by the Israel Science Foundation\\                                             
$^{e}$ &  supported by the Italian National Institute for Nuclear Physics (INFN) \\                
$^{f}$ &  supported by the Japanese Ministry of Education, Culture, Sports, Science and Technology 
          (MEXT) and its grants for Scientific Research\\                                          
$^{g}$ &  supported by the Korean Ministry of Education and Korea Science and Engineering          
          Foundation\\                                                                             
$^{h}$ &  supported by the Netherlands Foundation for Research on Matter (FOM)\\                   
$^{i}$ &  supported by the Polish State Committee for Scientific Research, project No.             
          DESY/256/2006 - 154/DES/2006/03\\                                                        
$^{j}$ &  partially supported by the German Federal Ministry for Education and Research (BMBF)\\   
$^{k}$ &  supported by RF Presidential grant N 1456.2008.2 for the leading                         
          scientific schools and by the Russian Ministry of Education and Science through its      
          grant for Scientific Research on High Energy Physics\\                                   
$^{l}$ &  supported by the Spanish Ministry of Education and Science through funds provided by     
          CICYT\\                                                                                  
$^{m}$ &  supported by the Science and Technology Facilities Council, UK\\                         
$^{n}$ &  supported by the US Department of Energy\\                                               
$^{o}$ &  supported by the US National Science Foundation. Any opinion,                            
findings and conclusions or recommendations expressed in this material                             
are those of the authors and do not necessarily reflect the views of the                           
National Science Foundation.\\                                                                     
$^{p}$ &  supported by the Polish Ministry of Science and Higher Education                         
as a scientific project (2009-2010)\\                                                              
$^{q}$ &  supported by FNRS and its associated funds (IISN and FRIA) and by an Inter-University    
          Attraction Poles Programme subsidised by the Belgian Federal Science Policy Office\\     
$^{r}$ &  supported by an FRGS grant from the Malaysian government\\                               
\end{tabular}                                                                                      
                                                           %
                                                           %

\pagenumbering{arabic} 
\pagestyle{plain}
\section{Introduction}
\label{sec-int}

 The production of multi-lepton final states in electron-proton 
 collisions\footnote{Here  and
 in the following, the term ``electron''  denotes generically both the
 electron ($e^-$) and the positron ($e^+$).} is predicted within the 
 framework of the Standard Model (SM). At  HERA energies,
 the production cross sections are small for high transverse momenta, $p_T$, 
 of the produced leptons and, along with the
 distributions of the kinematic quantities, can be calculated with high
 accuracy in the SM. Therefore contributions from
 beyond the SM could either be observed 
 as an increase of the visible cross sections or as a deviation from 
 the predicted distributions. 

 Multi-lepton final states were searched for 
 by the H1 Collaboration~\cite{pl:b668:268} using a luminosity of $463 \pbi$.
 The observed overall numbers of di- and tri-lepton events 
 were in good agreement with the SM predictions. However, some  
 events with large transverse momenta
 were observed, exceeding SM predictions in this region.

 The analysis presented here is based on  a luminosity of $480 \pbi$
 collected by the ZEUS experiment.
 Events with two or more high-$p_T$ leptons (electrons or muons)
 were searched for and the total yields and distributions of kinematic 
 variables were compared to SM predictions.
 In addition, the total visible and differential cross sections for di-lepton
 production were measured in the photoproduction regime, 
 in which the incoming electron has small squared
 momentum transfer, $Q^2<1 \gev^2$.

\section{Experimental set-up}
\label{sec-expandsim}

 The analysed  data  were  collected  between  1996  and  2007 at  the
 electron-proton collider HERA using  the ZEUS detector. During this
 period HERA operated with an  electron
 beam energy of $27.5\gev$
 and a proton  beam  energy of  $820\gev$ and, from
 1998,    of $920\gev$, corresponding   to centre-of-mass  energies of
 $300\gev$ and $318\gev$, respectively. 

\Zdetdesc

 Charged particles were tracked in the central tracking detector (CTD)~\citeCTD,
 which operated in a magnetic field of $1.43\Tesla$ provided by a thin
 superconducting solenoid and covered the polar-angle\footnote{
The ZEUS coordinate system is a right-handed Cartesian system,
with the $Z$ axis pointing in the proton beam direction, the $Y$
axis pointing up and the $X$ axis pointing towards the centre of HERA.
The polar angle, $\theta$, is measured with respect to the proton beam 
direction. The coordinate origin is at the nominal interaction point.}
region \mbox{$15^\circ<\theta<164^\circ$}. 
 Before the 2003--2007 running period, 
 the ZEUS tracking system was upgraded with a silicon microvertex detector
 (MVD) \cite{nim:a581:656}. The high-resolution uranium--scintillator 
 calorimeter (CAL)~\citeCAL consisted of three parts: the forward (FCAL), the barrel (BCAL)
 and the rear (RCAL) calorimeters. 
 The smallest subdivision of the CAL was called a cell. 
 The muon system consisted of rear, barrel (R/BMUON)~\cite{nim:a333:342}
 and forward (FMUON)~\cite{zeus:1993:bluebook}
 tracking detectors.
 The B/RMUON consisted of limited-streamer (LS) tube chambers
 placed behind the BCAL (RCAL), inside
 and outside a magnetised iron yoke
 surrounding the CAL.
 The barrel and rear muon chambers covered polar angles from $34^\circ$
 to $135^\circ$ and from $135^\circ$ to $171^\circ$, respectively.
 The FMUON consisted of six trigger planes of LS tubes
 and four planes of drift chambers
 covering the angular region from $5^\circ$ to $32^\circ$.
 The muon system exploited the magnetic field of the iron yoke and,
 in the forward direction, of two iron toroids magnetised to $\sim 1.6\Tesla$
 to provide a measurement of the muon momentum.

 The luminosity was measured using the Bethe-Heitler reaction 
 $ep\rightarrow e\gamma p$ by a luminosity detector which consisted of
 a lead--scintillator~\cite{desy-92-066,zfp:c63:391,acpp:b32:2025}  
 calorimeter
 and, in the 2003--07 running period, an independent magnetic 
 spectrometer~\cite{Helbich:2005qf}.
  The fractional systematic uncertainty 
  on the measured luminosity was 2.5\%.

 The integrated luminosity of the samples corresponds to 
 $480\pbi$ for events in which a search of electrons but no muons was carried out (electron channel)
 and to $444\pbi$ for events in which a search for either muons or electrons 
 was carried out (muon channel).
 The slight difference in integrated luminosity 
 is due to the requirement of a good performance of the 
 detector components involved in the search.

\section{Standard Model processes and Monte Carlo simulation}
\label{sec-simulation}

  To evaluate the detector acceptance and to provide simulations of 
  signal and 
  background distributions, Monte Carlo (MC) samples of signal and
  background events were generated.
  
  The SM predicts that isolated multi-lepton final states are 
 predominantly produced by two-photon interactions, 
 $\gamma \gamma \to  l^+ l^-$. 
The {\sc Grape} MC event generator \cite{cpc:136:126} was used
 to simulate these processes.
 It also includes contributions from $\gamma Z$ and $ZZ$ 
 interactions, photon internal conversions and virtual and real $Z$ 
 production. 
 It is based on the electroweak matrix elements at tree level.
 At the proton vertex, three contributions were considered:
 elastic, where the proton stays intact;
 quasi-elastic, where a resonant state is formed;
 and inelastic, where the proton interacts via its quark constituents.
 At the electron vertex, all values of  $Q^2$ were generated, from $Q^2 \simeq 0~\gev^2$  
 (photoproduction) to the deep inelastic scattering (DIS) regime.
 The uncertainty on the {\sc Grape} predictions was taken to be 
 3\%~\cite{pl:b668:268}.

 The Drell-Yan process from resolved photon events,
  in which the photon fluctuates into a $q\bar q$ pair,
 and the lepton pair is produced from the interaction between a quark in the proton and
 one of the quarks from the photon,
 is not included.
 However this is expected to be negligible in the investigated kinematic regime~\cite{ArteagaRomero:1991wn}.

The dominant SM background to topologies in which at least one
 electron is identified
comes from neutral current (NC) DIS and QED Compton (QEDC) events.
In NC ($ep \rightarrow eX$) events,  the scattered electron is 
identified as one of the electrons of the pair and
hadrons or photons in the hadronic system $X$ 
are misidentified as a further electron.
In QEDC events ($ep \rightarrow e\gamma X$),  the final-state photon may 
convert into an $e^+e^-$ pair in the detector material in front of the CTD and 
typically one of these two electrons is identified as the second electron of the pair. 

The NC DIS and QEDC events were simulated with the 
{\sc Djangoh}\cite{spi:www:djangoh11} and
{\sc Grape-Compton}~\cite{cpc:136:126} MC programs, respectively. 
The absolute predictions
of the {\sc Grape-Compton} MC were scaled  by a factor 1.13
in order to correct imperfections in the simulation of the dead material between  
the beampipe and the CTD. The uncertainty on this factor
was taken as a source of systematic uncertainty.
For the final-state topologies in which an electron and a muon were found, the background
from SM di-tau pair production was estimated using the {\sc Grape} MC program.

 Standard Model processes such as vector-meson
  ($\Upsilon$, charmonium) and open heavy-flavour (charm and beauty) 
  production were studied using the {\sc Diffvm}~\cite{proc:mc:1998:396} 
  and {\sc Pythia}~\cite{cpc:135:238} MC programs and were found to be negligible.

The generated events were passed through a full simulation of the ZEUS detector
based on the GEANT~\cite{tech:cern-dd-ee-84-1} program versions 3.13 (1996--2000) and 
3.21 (2003--07). They were then subjected to the same trigger requirements
and processed by the same reconstruction program as  the data.

\section{Event selection}
\label{sec-selection}

\subsection{Online selection}
\label{ssec-search-trigger}

Events with two or more leptons in the final state were selected using 
the ZEUS three-level trigger system  \cite{zeus:1993:bluebook,uproc:chep:1992:222,nim:a580:1257}.

To select electrons,
a significant energy deposit was required in the electromagnetic calorimeter
and at least one good track in the central detectors had to be present.
In addition, two other trigger chains
were used: the first,  dedicated to NC DIS selection, requiring the detection of
an electron with an energy $E^\prime_e> 4 \gev$; the second, dedicated to the 
selection of  
events with high transverse energy deposited 
in the calorimeter ($E_T > 25 \gev$).

To select muons~\cite{epj:c27:173},  a candidate was identified as 
a central track measured in the CTD  matched to  an energy 
deposit in the CAL and to a segment in the barrel or rear inner muon chambers.

\subsection{Electron identification}
\label{ssec-ofl-electron}
   
The following criteria were imposed to select electrons in the offline analysis:
\begin{itemize}
\item electron identification --- an algorithm~\cite{epj:c11:427}
 which combined information from the energy deposits in the
calorimeter and, when available, tracks measured in the central tracking
detectors was used to identify the electron candidates. 
Electron candidates in the central region
($20^\circ < \theta^e < 150^\circ$) were required to have  energy
greater than $10\gev$ and a track  matched with the energy
deposit in the calorimeter. The matched track was required 
to be fitted to the primary vertex and to have a momentum
of at least $3\gev$ and a distance of closest approach between the energy 
deposition and the track of less than $8$ cm. 
Forward electrons
($5^\circ < \theta^e < 20^\circ$) were also required to have an energy greater
than $10\gev$ while, for electrons in the rear 
region ($150^\circ < \theta^e < 175^\circ$), the energy requirement 
was decreased to $5\gev$;
\item isolation ---
to ensure high purity, each electron candidate was required
to be isolated such that the total energy not associated with the
electron in an $\eta - \phi$ cone of radius 0.8 centred on the electron
was less than $0.3\gev$. 
This requirement was complemented, for electrons in the central region,
 by the request that no track with $p_T> 1\gev$,
 other than the matching track,  was contained
in an $\eta - \phi$ cone of radius 0.4 centred on the electron;
\item QEDC background reduction --- for the data collected in 2003--07, each track associated with an electron candidate was required to have at least two hits in the MVD.
 This requirement removed photon conversions in the material between the MVD and  the CTD.
\end{itemize}

\subsection{Muon identification}
\label{ssec-ofl-muon}

The following criteria were imposed to select muons in the offline analysis:
\begin{itemize}
\item muon identification ---
 at least one muon candidate in the event was required to be reconstructed by the rear, 
 barrel or forward muon chambers, matched to a track and to an energy deposit in the 
 calorimeter. In the case when only one muon in the event was reconstructed by the muon 
 chambers, additional muons were also selected with looser criteria, by requiring a track pointing 
 towards a calorimeter energy deposit compatible with that from a minimum ionising particle (mip). 
 Each muon candidate was required to be associated with a track fitted to the primary vertex.
 The muon momentum was reconstructed using the central 
 tracking devices, complemented with the
 information from the FMUON when available.   
 The muon was required to have $p_T^{\mu}>2\gev$, and to lie
        in the angular region $20^\circ < \theta^\mu < 26^\circ$ 
        (FMUON), 
        $35^\circ <\theta^\mu<160^\circ$ (B/RMUON), 
         $20^\circ < \theta^\mu < 160^\circ$ (mip);
 \item isolation --- to ensure high purity, each identified muon was 
  required to be isolated such that only the matching track was 
  contained in an $\eta - \phi$  cone of radius 1.0
  centred on the muon.  This cut, harder than in the electron selection, was used 
   to reject background  events in which a muon was found very close to a hadronic system, 
  in particular in the $e\mu$ channel;
 \item cosmic-muon background reduction --- the reconstructed primary vertex had to be consistent
 with the HERA beam-spot position. If two muons were found, the acollinearity angle, $\Omega$, between the two muons had to satisfy \mbox{$\cos\Omega > -0.995$}. For events with 
 \mbox{$\cos\Omega < -0.990$}, additional CAL timing cuts were applied. 
\end{itemize}

\subsection{Event selection and classification}
\label{ssec-ofl-final}

The final event selection required  the event vertex to be reconstructed
with $|Z_{\rm VTX}| < 30 \cm$.
At least two leptons, electrons  or muons, had 
to   be reconstructed  in the  central    part of  the detector
($20^\circ  < \theta^l  <  150^\circ$).
 One  of the  leptons had to have
$p_T^{l_1}  > 10 \gev$ and  the other  $p_T^{l_2}>5 \gev$. Additional leptons
identified as described in Sections
\ref{ssec-ofl-electron} and \ref{ssec-ofl-muon} could be present in the event.
No explicit requirement on the charge of the leptons was imposed.
According to the   number  and the
flavour  of  the lepton candidates,  the   events were classified into
mutually exclusive samples.

For the measurement of the production cross section of 
$e^+e^-$ and $\mu^+\mu^-$ pairs in the
photoproduction regime,  
the cut  $(E-P_Z)<45\gev$ was applied. This quantity was reconstructed in the electron case as 
\begin{equation}
E-P_Z = \sum_i E^{\rm corr}_i(1-\cos(\theta_i)),
\end{equation}
where 
the sum runs over the corrected energies, $E_i^{\rm corr}$, of the CAL
clusters
and, in the muon case, as
\begin{equation}
E-P_Z = \sum_i E_i(1-\cos(\theta_i))-\sum_{\rm mip}E_{\rm mip}(1-\cos(\theta_{\rm mip}))+
\sum_{\rm muon}E_{\rm muon}(1-\cos(\theta_{\rm muon})),
\end{equation}
where $E_i$ is the energy of the $i^{\rm th}$ CAL cell and 
 the $(E-P_Z)$ of the CAL mip was replaced by that of the muon track.
This requirement selects events in which the  scattered electron was lost in
the beampipe and corresponds to a cut of $Q^2<1\gev^2$ and
 on the event inelasticity, $y=(E-P_Z)/2E_e<0.82$, 
where $E_e$ is the electron beam energy.
The background from NC DIS and QEDC events is negligible in this sample, which will be referred
to as the $\gamma\gamma$ sample in the following.

\section{Systematic uncertainties}
\label{sec-syst}

The following sources of systematic uncertainties were considered; 
the effect on the total visible cross section is given:
\begin{itemize}
\item the muon acceptance, including the B/RMUON trigger, the reconstruction 
         and  the muon identification efficiencies, is known to about 7\% from a 
         study based on an independent elastic di-muon sample \cite{thesis:turcato:2002}, 
         resulting in an uncertainty of ($^{+10\%}_{-8\%}$) for muons;
 \item the uncertainty on the efficiency of the CTD part of the trigger chain was estimated from a study based
 	on an independent sample of low-multiplicity low-$Q^2$ DIS events~\cite{thesis:marchesini:2007},
	resulting in an uncertainty of +5\% for electrons and $\pm 5\%$ for muons; 
  \item the CAL energy scale was varied by its uncertainty of 3\%, resulting in an uncertainty 
  of  ($^{+4\%}_{-3\%}$) for electrons and negligible for muons;
\item the uncertainty on the efficiency of the CAL part of the muon trigger 
($\pm 3\%$) and of the mip finder ($\pm 2\%$)  resulted 
in an uncertainty of  $\pm 4\%$ for muons; 
\item the uncertainty on the measurement of the hadronic system was evaluated by using an alternative reconstruction of $E-P_Z$,
 resulting in an uncertainty of $-1.8\%$ for electrons; 
  \item the scaling factor of the  QEDC  MC was varied between 0.95 and 1.31, 
  as allowed by the comparison with a QEDC-enriched data sample,
  resulting in a negligible effect for both electrons and muons.
\end{itemize}

The total systematic uncertainty was obtained by adding the individual
contributions in quadrature. A 2.5\% overall normalisation uncertainty
associated with the luminosity measurement was included only in the systematic
uncertainty of the total visible cross section.

\section{Results}
\label{sec-xsec}

 The number of selected events in the data are compared to SM predictions
 in \tab{comparison}.  The following different di- and tri-lepton topologies are
 listed: $ee$, $\mu\mu$, $e \mu$, $eee$ and $e \mu\mu$.
  The observed number of events is in good agreement with the predictions of the SM,
 according to which the  NC DIS and
 QEDC processes give  a sizeable contribution to the  $ee$  channel. 
 Most of the events contributing to the $e\mu$ topology are predicted to  come from
 di-muon production at high $Q^2$, in which the beam electron is scattered
 at large angles and is therefore seen in the detector, while one of the
 muons  is outside the acceptance region. A small contribution
 to this channel ($\sim 2$ events) is predicted to  
 come from di-$\tau$ production,
 while the NC DIS background constitutes $\sim 10\%$ of the sample.
 
 Three four-lepton events, 2  in the $ee\mu\mu$ 
 and 1 in the $eeee$ channel, were observed, to be compared to a SM expectation of 
 $\sim 1$. 
 The contributions from true four-lepton events are not 
 included in the SM predictions and are expected to be small.
 Events with other multi-lepton topologies were searched for, but none was found. 

 In Tables~\ref{tab-comparison} and \ref{tab-comparison_m100} the NC DIS and QEDC background contributions are given
 as limits
 at 95\% confidence level (C.L.) when none or few events were selected from
 the background MC samples. When this is done, the central value of the 
 total SM prediction is determined as the most probable value (mode) of the 
 convolution of the Gaussian signal
 distribution with the poissonian background distributions, and the 
 uncertainty on the total SM prediction is determined by taking the 68\% C.L. interval.

\begin{figure}
\begin{center}
  \begin{picture}(594,568)(-65,0)
    \put(0,0){\includegraphics[width=11cm]{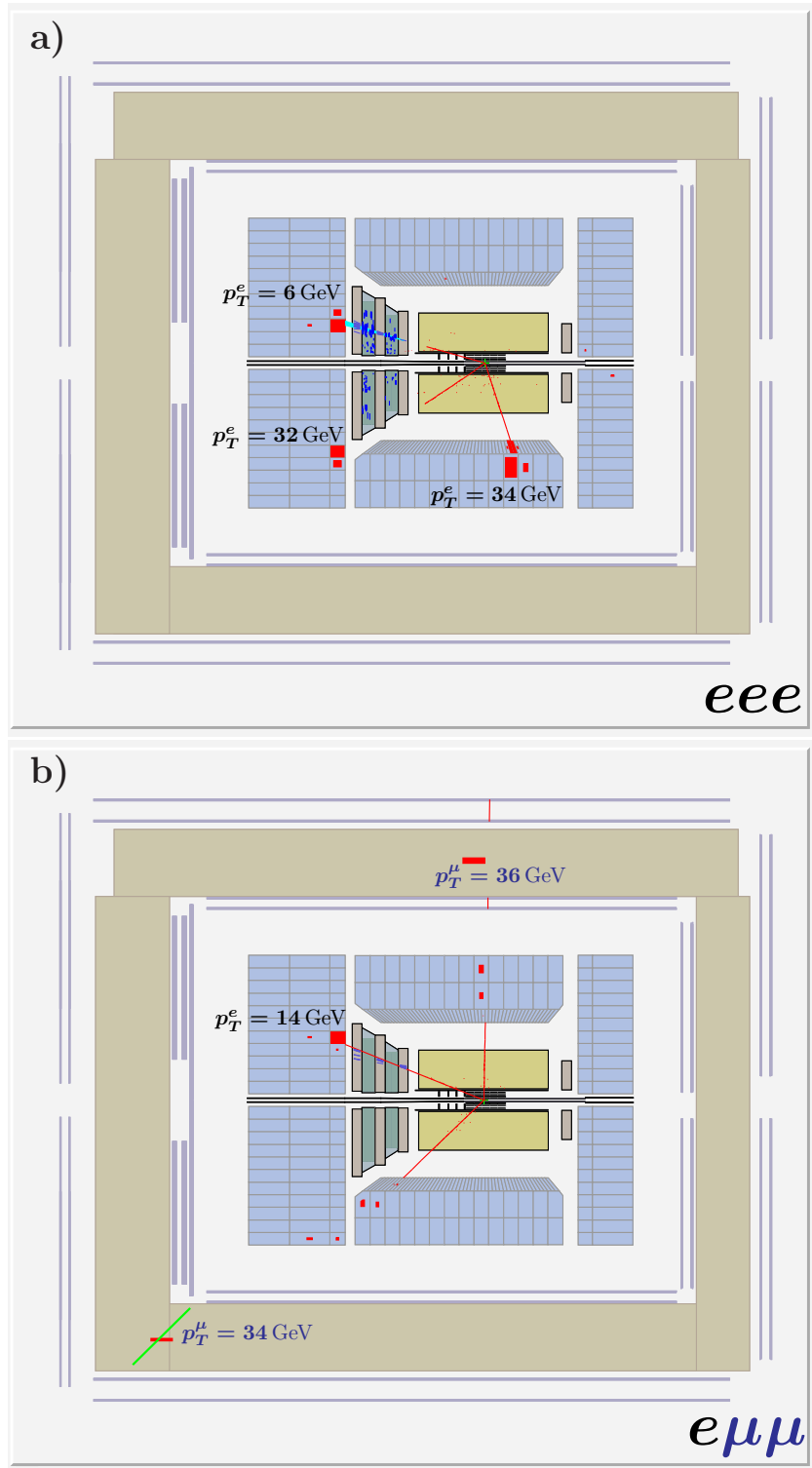}}
    \put(10,555){\bf \large a)}
    \put(10,270){\bf \large b)}
  \end{picture}
\end{center}
\caption{a) An event with three electron candidates in the ZEUS detector.
 The invariant mass of the two highest-$p_T$
 electrons is $M_{12} =113\gev$; the corresponding transverse momenta are
 given above.
 b) An event with two muons and an electron candidate 
 in the ZEUS detector.
 The invariant mass of the di-muon pair is $77.5\gev$; the corresponding 
 transverse momenta are given above.
}
\label{fig-evento}
\end{figure}

 Two events, one with three electrons in the final state and
  one with two muons 
 and an electron, passing the three-lepton selection cuts, are shown in 
 Fig.~\ref{fig-evento}.
 \subsection{Kinematic distributions}
\begin{figure}
\begin{center}
  \begin{picture}(594,481)(0,0)
    \put(0,0){\includegraphics[width=17cm]{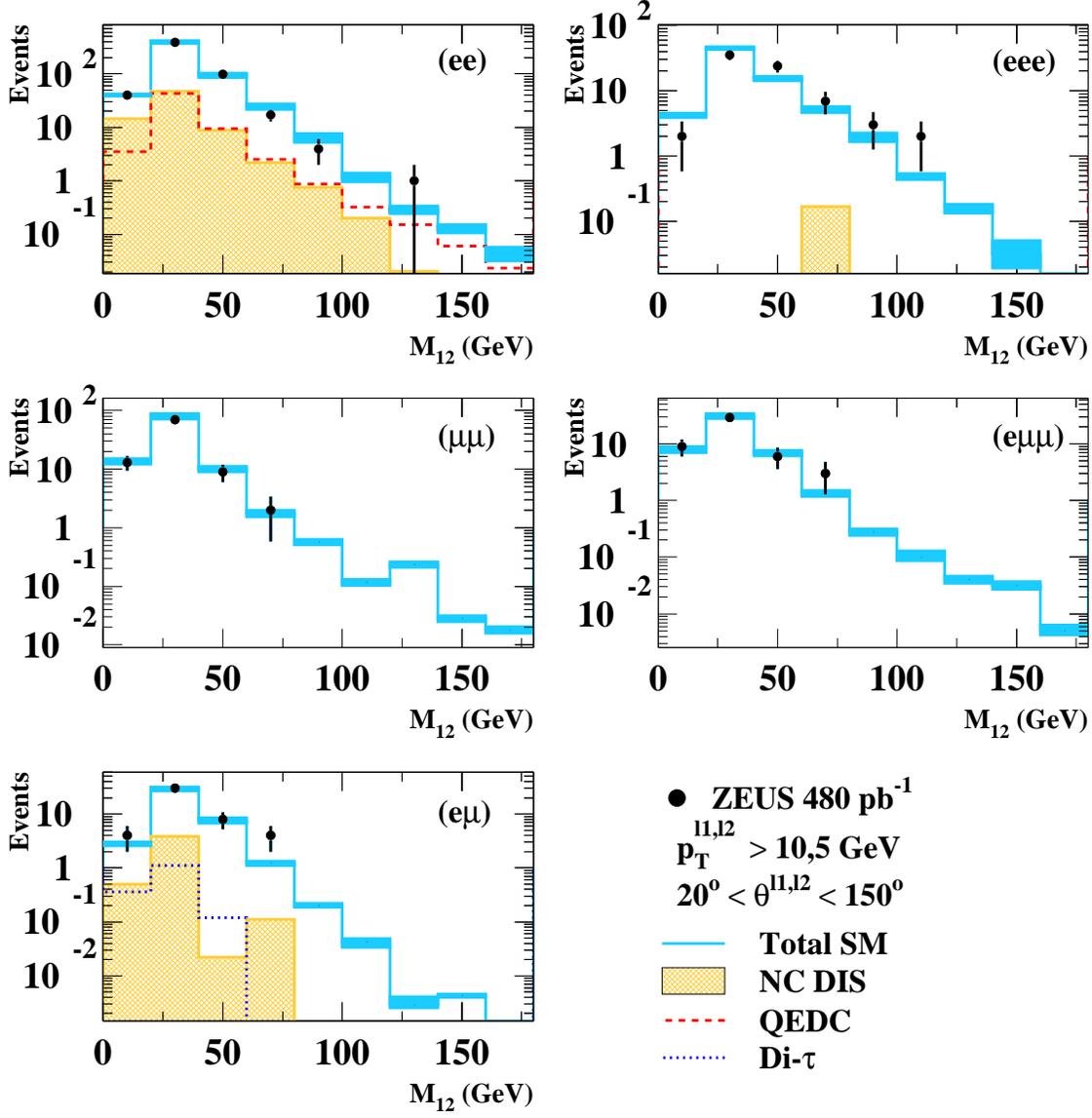}}
  \end{picture}
\end{center}
\caption{Distributions of the invariant mass of the two highest-$p_T$ leptons for the different multi-lepton topologies: $ee$, $eee$, $\mu\mu$, $e\mu\mu $, $e \mu $.  
  The ZEUS data are displayed as the full dots. 
  The errors on the data are given by the square
  root of the number of events in each bin.
  The SM predictions are represented as the solid line and are obtained by 
  summing the contributions of
  di-lepton production, NC DIS, QED Compton events, and, for the $e\mu$
  channel, di-tau production. The error band represents the systematic 
  uncertainty on the SM predictions.
  }
\label{fig:distributions_mass}
\end{figure}

\begin{figure}
\begin{center}
  \begin{picture}(594,481)(0,0)
    \put(0,0){\includegraphics[width=17cm]{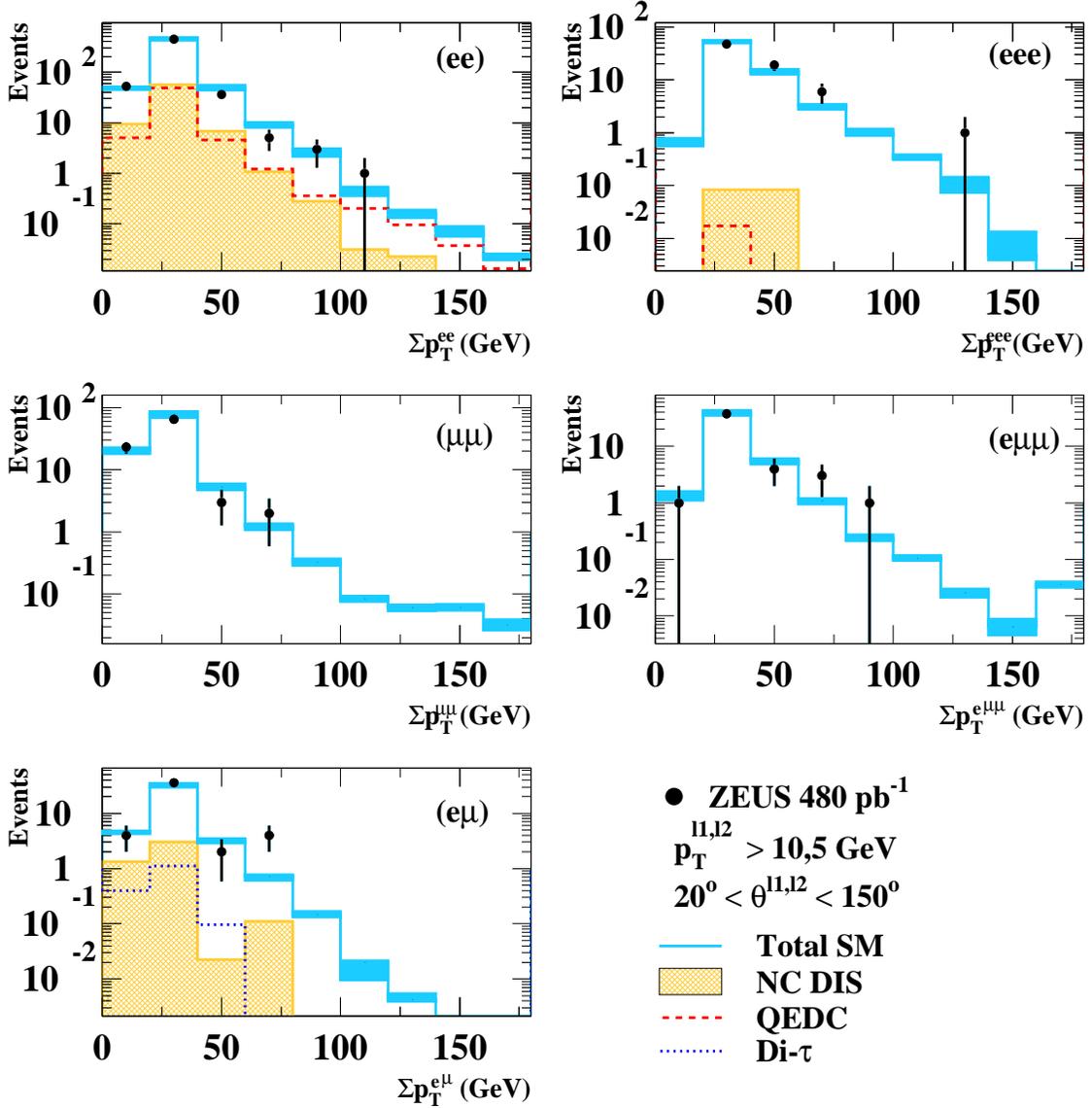}}
  \end{picture}
\end{center}
\caption{Distributions of the sum of the transverse momenta of the leptons for the different multi-lepton topologies: $ee$, $eee$, $\mu\mu$, $e \mu\mu$, $e\mu$. 
  The ZEUS data are displayed as the full dots. 
  The errors on the data are given by the square
  root of the number of events in each bin.
  The SM predictions are represented as the solid line and are obtained by 
  summing the contributions of
  di-lepton production, NC DIS, QED Compton events, and, for the $e\mu$
  channel, di-tau production. The error band represents the systematic 
  uncertainty on the SM predictions.
}
\label{fig:distributions_sumpt}
\end{figure}

 The distributions of the mass of the two highest-$p_T$ leptons in the
 event, $M_{12}$,  and of the scalar  sum of the transverse momenta of
 all the identified  leptons in the  event, $\sum{p_T^l}$,  are shown in Figs.~\ref{fig:distributions_mass}
 and \ref{fig:distributions_sumpt} for all the observed
 di- and tri-lepton topologies, and are compared to SM predictions. 
 The SM gives a good description of the data. 
 In the  mass region between 80 and 100\gev, which is sensitive to $Z^0$ production,
 7 events were observed in the data, 
 compatible with the predictions from the SM of $\sim 9$ 
 events, including $\sim 1$ event from real $Z^0$ production.

 The high-mass and high-$\sum{p_T^l}$ regions are particularly
 sensitive to possible contributions from physics beyond the SM. 
 The event yields for $M_{12}> 100\gev$
 for all the observed di- and tri-lepton 
 channels are summarised in \tab{comparison_m100}. In the electron channels,
 3 events at high masses are observed, to be compared with a SM prediction of 
 2.5. Two of these events are observed in the $eee$ topology, for which the 
 SM expectation is 0.7.  No event with $M_{12}> 100~\gev$  is seen in the muon channels. 
 The event yield for $\sum{p_T^l}>100 \gev$, combined for all the lepton topologies, 
 is summarised in \tab{comparison_sumpt100}.
 Two events at high-$\sum{p_T^l}$ are observed, to be compared with a SM prediction
 of $\sim 1.6$.

\begin{figure}
\begin{center}
  \begin{picture}(594,298)(0,0)
    \put(0,0){\includegraphics[width=17cm]{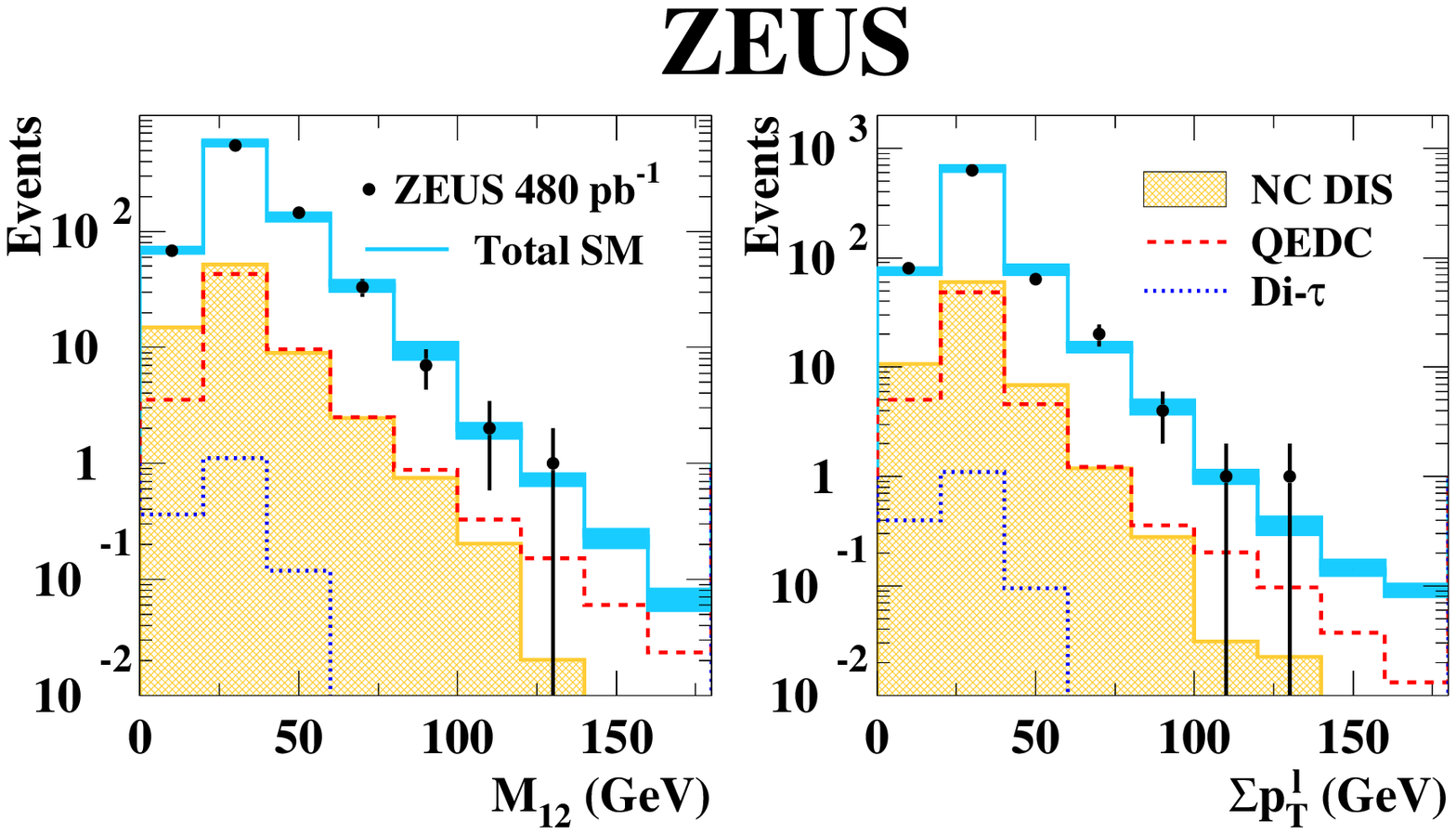}}
    \put(60, 45){\bf \large (a)}
    \put(275,45){\bf \large (b)}
  \end{picture}
\end{center}
\caption{Distributions of (a) the invariant mass of the two highest-$p_T$ leptons and (b) the sum of the transverse momenta of the leptons for all the individual lepton topologies combined: $ee$, $eee$, $\mu\mu$, $e\mu\mu$, $e\mu$.  
  The ZEUS data are displayed as the full dots. 
  The errors on the data are given by the square
  root of the number of events in each bin.
  The SM predictions are represented as the solid line and are obtained by 
  summing the contributions of
  di-lepton production, NC DIS, QED Compton events, and, for the $e\mu$
  channel, di-tau production. The error band represents the systematic 
  uncertainty on the SM predictions.
}
\label{fig:distributions_comb}
\end{figure}

 The distributions of $M_{12}$ and $\sum{p_T^l}$, combined for all the di- and 
 tri-lepton topologies, are shown in Fig.~\ref{fig:distributions_comb}. Also
 in this case, the data are well described by the SM predictions.

\subsection{Cross sections}
 Total visible and differential cross sections for di-electron and di-muon 
 production were
 determined  in the kinematic region defined by: 
\begin{center} 
$p_T^{l_1} > 10 \gev$,  $p_T^{l_2} > 5 \gev$, 
\mbox{$20^\circ < \theta^{l_{1,2}} < 150^\circ$}, $Q^2<1 \gev^2$, 
$y<0.82$.
\end{center}
 The cross sections are given at $\sqrt{s}=318 \gev$: the small ($\sim 5\%$)
 correction
 needed for the 1996--97 data sample was extracted from the MC. 
 The effect of final-state radiation on the cross section was checked and found to
 be negligible.

The total visible cross sections, corrected for acceptance, were measured to be
 \begin{eqnarray}
 \sigma(\gamma\gamma \rightarrow e^+e^-) & = & 0.64\pm 0.05^{+0.04}_{-0.03}~{\rm pb}
\end{eqnarray}
for the electron channel, and
 \begin{eqnarray}
\sigma(\gamma\gamma \rightarrow \mu^+\mu^-) & = &  0.58\pm 0.07^{+0.07}_{-0.06}~{\rm pb}
\end{eqnarray}
for the muon channel. 

 Since the muon and electron cross sections differ only marginally, they 
 were combined in a single measurement, evaluated as the 
 weighted mean of the two \cite{Demortier:1999vv},
 assuming the systematic uncertainties to be uncorrelated.
 The systematic uncertainties of each measurement were
 symmetrised before the combination, by taking as systematic uncertainty the
 largest between the negative and the positive.
 The total visible cross sections are shown in \tab{xsec},
 compared with the SM predictions.

\begin{figure}
\begin{center}
  \begin{picture}(594,481)(0,0)
    \put(0,0){\includegraphics[height=17cm]{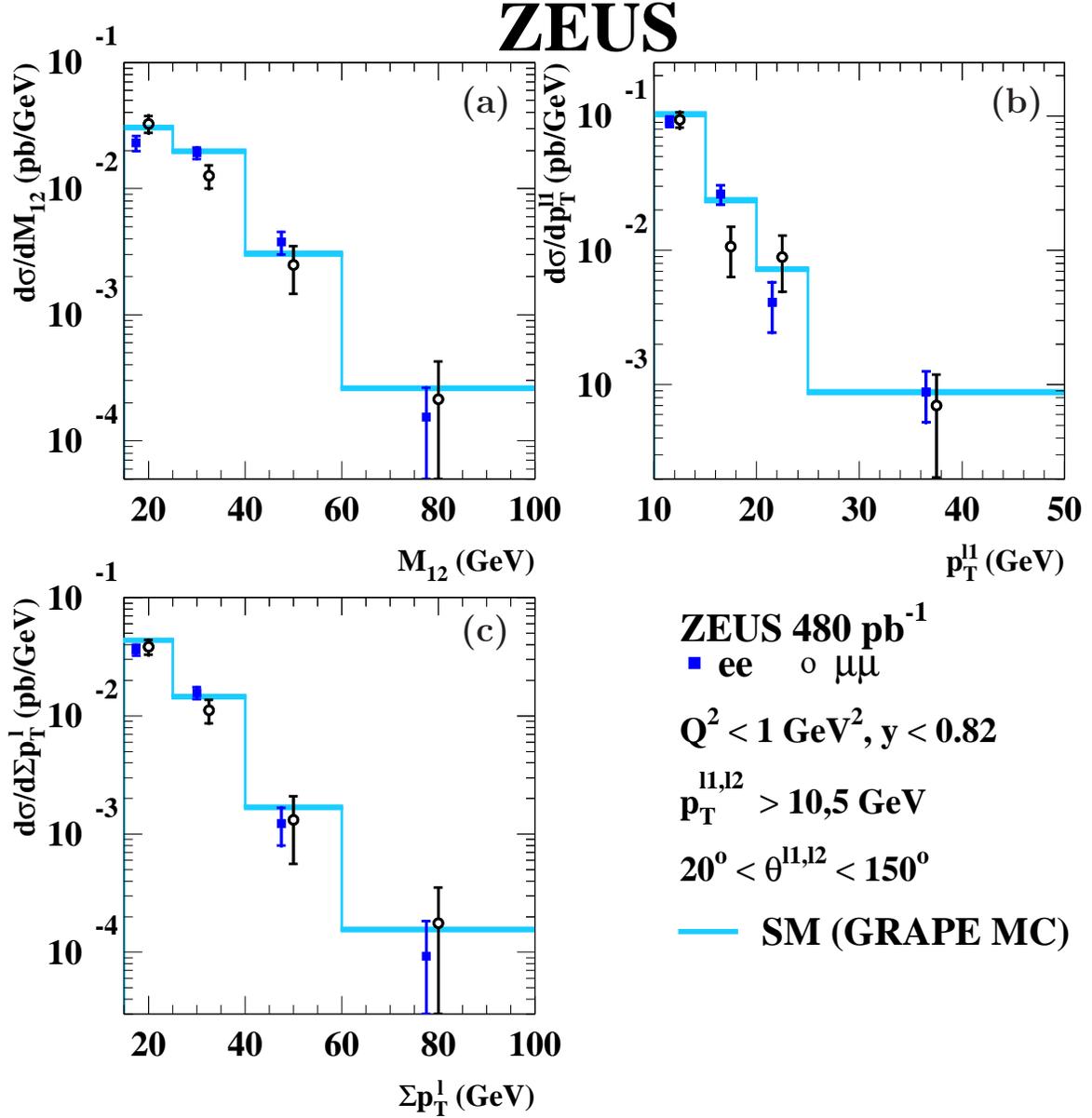}}
    \put(190,415){\bf \large (a)}
    \put(405,415){\bf \large (b)}
    \put(190,200){\bf \large (c)}
  \end{picture}
\end{center}
\caption{Differential cross sections as a function of (a) the invariant mass of
 the lepton pair, $M_{12}$, (b) the transverse momentum $p_T^{\l_{1}}$ of the highest-$p_T$ lepton, and (c) the scalar sum of the transverse momenta of the 
two  highest-$p_T$ leptons, $\sum{p_T^l}$. The di-muon cross sections are 
shown as the open dots, while the full dots are the di-electron measurements, which have been displaced for clarity.
 The data are compared with the predictions of the
 {\sc Grape} Monte Carlo. The full error bars are the quadratic sum of
 the statistical (inner part) and systematic uncertainties.}
\label{fig-xsec}
\end{figure}

 Differential cross sections as a function of the invariant mass, $M_{12}$,
 the transverse momentum of the highest-$p_T$ lepton, $p_T^{l_1}$, and the
 scalar sum of the transverse momentum of the two leptons, $\sum{p_T^l}$,
  are shown in Fig.~\ref{fig-xsec},  separately 
 for electrons and muons. 
 The di-electron, di-muon and combined cross sections
 are summarised in \tab{xsec2}.
 The combination was done as described for the total visible cross section.
 Good agreement is observed between the data and the SM predictions.

\section{Conclusions}
\label{sec-conclusion}
 Events with two or more isolated leptons with high transverse momentum
 were observed using the full data sample taken with the ZEUS detector at HERA.
 The total number of multi-lepton events for different lepton configurations
 as well as their $p_T$ and mass distributions were studied. No significant 
 deviations from the predictions of the SM were observed. 
 In addition, the total visible and differential cross sections 
 for the $e^+e^-$ and $\mu^+\mu^-$ signatures were measured in 
 photoproduction and were observed to be in good agreement 
 with the SM predictions.


\section{Acknowledgements}

We appreciate the contributions to the construction and maintenance of the 
ZEUS detector of many people who are not listed as authors. The HERA machine 
group and the DESY computing staff are especially acknowledged for their
success in providing excellent operation of the collider and the data-analysis
environment. We thank the DESY directorate for their strong support and 
encouragement.

\vfill\eject

{
\def\bibname{\Large\bf References}
\def\refname{\Large\bf References}
\pagestyle{plain}
\ifzeusbst
  \bibliographystyle{./BiBTeX/bst/l4z_default}
\fi
\ifzdrftbst
  \bibliographystyle{./BiBTeX/bst/l4z_draft}
\fi
\ifzbstepj
  \bibliographystyle{./BiBTeX/bst/l4z_epj}
\fi
\ifzbstnp
  \bibliographystyle{./BiBTeX/bst/l4z_np}
\fi
\ifzbstpl
  \bibliographystyle{./BiBTeX/bst/l4z_pl}
\fi
{\raggedright
\bibliography{./BiBTeX/user/syn,%
              ./BiBTeX/bib/l4z_articles,%
              ./BiBTeX/bib/l4z_books,%
              ./BiBTeX/bib/l4z_conferences,%
              ./BiBTeX/bib/l4z_h1,%
              ./BiBTeX/bib/l4z_misc,%
              ./BiBTeX/bib/l4z_old,%
              ./BiBTeX/bib/l4z_preprints,%
              ./BiBTeX/bib/l4z_replaced,%
              ./BiBTeX/bib/l4z_temporary,%
              ./BiBTeX/bib/l4z_zeus}}

\providecommand{\etal}{et al.\xspace}
\providecommand{\coll}{Collaboration}
\catcode`\@=11
\def\@bibitem#1{%
\ifmc@bstsupport
  \mc@iftail{#1}%
    {;\newline\ignorespaces}%
    {\ifmc@first\else.\fi\orig@bibitem{#1}}
  \mc@firstfalse
\else
  \mc@iftail{#1}%
    {\ignorespaces}%
    {\orig@bibitem{#1}}%
\fi}%
\catcode`\@=12
\begin{mcbibliography}{10}

\bibitem{pl:b668:268}
H1 \coll, F.D.~Aaron \etal,
\newblock Phys.\ Lett.{} B~668~(2008)~268\relax
\relax
\bibitem{zeus:1993:bluebook}
ZEUS \coll, U.~Holm~(ed.),
\newblock {\em The {ZEUS} Detector}.
\newblock Status Report (unpublished), DESY (1993),
\newblock available on
  \texttt{http://www-zeus.desy.de/bluebook/bluebook.html}\relax
\relax
\bibitem{nim:a279:290}
N.~Harnew \etal,
\newblock Nucl.\ Inst.\ Meth.{} A~279~(1989)~290\relax
\relax
\bibitem{npps:b32:181}
B.~Foster \etal,
\newblock Nucl.\ Phys.\ Proc.\ Suppl.{} B~32~(1993)~181\relax
\relax
\bibitem{nim:a338:254}
B.~Foster \etal,
\newblock Nucl.\ Inst.\ Meth.{} A~338~(1994)~254\relax
\relax
\bibitem{nim:a581:656}
A.~Polini et al.,
\newblock Nucl.\ Inst.\ Meth.{} A~581~(2007)~31\relax
\relax
\bibitem{nim:a309:77}
M.~Derrick \etal,
\newblock Nucl.\ Inst.\ Meth.{} A~309~(1991)~77\relax
\relax
\bibitem{nim:a309:101}
A.~Andresen \etal,
\newblock Nucl.\ Inst.\ Meth.{} A~309~(1991)~101\relax
\relax
\bibitem{nim:a321:356}
A.~Caldwell \etal,
\newblock Nucl.\ Inst.\ Meth.{} A~321~(1992)~356\relax
\relax
\bibitem{nim:a336:23}
A.~Bernstein \etal,
\newblock Nucl.\ Inst.\ Meth.{} A~336~(1993)~23\relax
\relax
\bibitem{nim:a333:342}
G.~Abbiendi \etal,
\newblock Nucl.\ Inst.\ Meth.{} A~333~(1993)~342\relax
\relax
\bibitem{desy-92-066}
J.~Andruszk\'ow \etal,
\newblock Preprint \mbox{DESY-92-066}, DESY, 1992\relax
\relax
\bibitem{zfp:c63:391}
ZEUS \coll, M.~Derrick \etal,
\newblock Z.\ Phys.{} C~63~(1994)~391\relax
\relax
\bibitem{acpp:b32:2025}
J.~Andruszk\'ow \etal,
\newblock Acta Phys.\ Pol.{} B~32~(2001)~2025\relax
\relax
\bibitem{Helbich:2005qf}
M.~Helbich \etal,
\newblock Nucl.\ Inst.\ Meth.{} A~565~(2006)~572\relax
\relax
\bibitem{cpc:136:126}
T.~Abe,
\newblock Comp.\ Phys.\ Comm.{} 136~(2001)~126\relax
\relax
\bibitem{ArteagaRomero:1991wn}
N.~Arteaga-Romero, C.~Carimalo and P.~Kessler,
\newblock Z. Phys.{} C~52~(1991)~289\relax
\relax
\bibitem{spi:www:djangoh11}
H.~Spiesberger,
\newblock {\em {\sc heracles} and {\sc djangoh}: Event Generation for $ep$
  Interactions at {HERA} Including Radiative Processes}, 1998,
\newblock available on \texttt{http://www.desy.de/\til
  hspiesb/djangoh.html}\relax
\relax
\bibitem{proc:mc:1998:396}
B.~List and A.~Mastroberardino,
\newblock {\em Proc.\ Workshop on Monte Carlo Generators for {HERA} Physics},
  p.~396.
\newblock DESY, Hamburg, Germany (1999).
\newblock Also in preprint \mbox{DESY-PROC-1999-02},
\newblock available on \texttt{http://www.desy.de/\til heramc/}\relax
\relax
\bibitem{cpc:135:238}
T.~Sj\"{o}strand \etal,
\newblock Comp.\ Phys.\ Comm.{} 135~(2001)~238\relax
\relax
\bibitem{tech:cern-dd-ee-84-1}
R.~Brun et al.,
\newblock {\em {\sc geant3}},
\newblock Technical Report CERN-DD/EE/84-1, CERN, 1987\relax
\relax
\bibitem{uproc:chep:1992:222}
W.H.~Smith, K.~Tokushuku and L.W.~Wiggers,
\newblock {\em Proc.\ Computing in High-Energy Physics (CHEP), Annecy, France,
  Sept. 1992}, C.~Verkerk and W.~Wojcik~(eds.), p.~222.
\newblock CERN, Geneva, Switzerland (1992).
\newblock Also in preprint \mbox{DESY 92-150B}\relax
\relax
\bibitem{nim:a580:1257}
P.D.~Allfrey \etal,
\newblock Nucl.\ Inst.\ Meth.{} A~580~(2007)~1257\relax
\relax
\bibitem{epj:c27:173}
ZEUS \coll, S.~Chekanov \etal,
\newblock Eur.\ Phys.\ J.{} C~27~(2003)~173\relax
\relax
\bibitem{epj:c11:427}
ZEUS \coll, J.~Breitweg \etal,
\newblock Eur.\ Phys.\ J.{} C~11~(1999)~427\relax
\relax
\bibitem{thesis:turcato:2002}
M.~Turcato.
\newblock Ph.D.\ Thesis, Universit\`a degli Studi di Padova, Italy, Report
  \mbox{DESY-THESIS-03-039}, 2002\relax
\relax
\bibitem{thesis:marchesini:2007}
I.~Marchesini.
\newblock Diploma \ Thesis, Universit\`a degli Studi di Padova, Italy, 2007
  (unpublished)\relax
\relax
\bibitem{Demortier:1999vv}
CDF and \DO \coll, L.~Demortier \etal,
\newblock FERMILAB-TM-2084{}~(1999)\relax
\relax
\end{mcbibliography}
}
\vfill\eject



\begin{table}
  \begin{center}
     {\bf ZEUS} {\boldmath(${\mathcal L} =  480 \pbi$)}
    \begin{tabular}{|c|c|c||c|c|c|}
      \hline
      Topology & Data  & Total SM & Multi-lepton Production & NC DIS & Compton \\
      \hline
$ee$           & 545 & $563^{+29}_{-37}$ & $429^{+21}_{-29}$ & $74\pm 5$ & $60\pm 10$  \\
$\mu\mu$       & 93 & 106$\pm$12 & 106$\pm$12 & $<0.5$ & $-$ \\
$e\mu$         & 46 & 42$\pm$4 & $37^{+3}_{-4}$  & 4.5$\pm$1.2 & $-$  \\ 
$eee$          & 73 & $75^{+5}_{-4}$ &   $73^{+4}_{-5}$ & $<1$ & $<3$  \\
$e\mu\mu$      & 47 & 48$\pm$5 & 48$\pm$5      & $<0.5$ & $-$ \\
$eeee$         & 1  & $0.9^{+0.5}_{-0.1}$ & 0.6$\pm$0.1 & $< 0.4$ & $< 1$ \\
$ee\mu\mu$    & 2  & $0.5^{+0.3}_{-0.1}$ & 0.4$\pm$0.1 & $<0.5$ & $-$ \\
      \hline
All 4 leptons  & 3 & $1.4^{+0.7}_{-0.1}$ & $1.0\pm 0.2$ &  \multicolumn{2}{c|}{$<1.4$} \\ \hline
$ee$ ($\gamma\gamma$ sample)     & 166 & $185^{+~8}_{-14}$ & $183^{+~8}_{-14}$ & 1.4$\pm$1.0 & $1.4\pm 0.6$ \\
$\mu\mu$ ($\gamma\gamma$ sample) & 72  & $85^{+~9}_{-10}$  & $85^{+~9}_{-10}$  & $<0.5$ & $-$ \\
      \hline
      \hline
    \end{tabular}
  \end{center}
  \caption{The observed and predicted multi-lepton event yields for
           the $ee$, $\mu\mu$, $e\mu$, $eee$, $e\mu\mu$, 
           $eeee$ and $ee\mu\mu$ event topologies;
           the event yields for the $ee$ and $\mu\mu$ topologies in the 
           $\gamma \gamma$ samples.
           The quoted uncertainties consist of model uncertainties,
           MC statistical uncertainties and 
           systematic experimental uncertainties added in quadrature.
	   Limits at 95\% C.L. are given when none or few events were selected
           from the background MC samples. The central value and the 
           uncertainty on the total SM predictions
           are in these cases determined as explained in the text.
}
  \label{tab-comparison}
\end{table}

\begin{table}
  \begin{center}
     {\bf ZEUS} {\boldmath(${\mathcal L} = 480 \pbi$)}
    \begin{tabular}{|c|c|c||c|c|c|}
      \hline
      Topology, & \multirow{2}{*}{Data}  & \multirow{2}{*}{Total SM} & 
                  \multirow{2}{*}{Multi-lepton Production} & 
                  \multirow{2}{*}{NC DIS} & 
                  \multirow{2}{*}{Compton} \\
      $M_{12}>100\gev$ & & & & & \\ \hline
      $ee$  & 1 & $1.7 \pm 0.2$ & $0.9\pm 0.1$ & $0.2\pm 0.1$ & $0.6\pm 0.1$ \\
      $\mu\mu$  & 0 & $0.4\pm0.1$ & $0.4\pm 0.1$ & $<0.01$ & $-$ \\
      $e\mu$    & 0 &  $0.06^{+0.03}_{-0.01}$ & $0.05\pm 0.02$ & $<0.02$ & $-$ \\
      $eee$     & 2 & $0.7\pm 0.1$ & $0.7\pm 0.1$ & $<0.01$ & $<0.02$ \\
      $e\mu\mu$ & 0 & $0.18\pm 0.05$ & $0.18\pm 0.05$ & $<0.01$ & $-$ \\
\hline
    \end{tabular}
  \end{center}
  \caption{The observed and predicted  high-mass, $M_{12}>100\gev$,
           multi-lepton event yields.
	   The invariant mass 
           was calculated using the two highest-$p_T$ leptons.
           The quoted uncertainties consist of model uncertainties,
           MC statistical uncertainties and 
           systematic experimental uncertainties added in quadrature.
	   Limits at 95\% C.L. are given when none or few events were selected
           from the background MC samples. The central value and the 
           uncertainty on the total SM predictions
           are in these cases determined as explained in the text.
}
  \label{tab-comparison_m100}
\end{table}
\begin{table}
  \begin{center}
     {\bf ZEUS} {\boldmath(${\mathcal L} = 480 \pbi$)}
    \begin{tabular}{|c|c||c|c|c|}
      \hline
      \multicolumn{5}{|c|}{$\sum{p_T^l}>100\gev$} \\ \hline
      Data  & {Total SM} & {Multi-lepton Production} & {NC DIS} & {Compton} \\ \hline
       2  & $1.56 \pm 0.15$ & $1.16\pm 0.13$ & $0.05\pm 0.02$ & $0.35\pm 0.06$ \\
\hline
    \end{tabular}
  \end{center}
  \caption{The observed and predicted high-$\sum{p_T^l}$
           multi-lepton event yields for all topologies combined, where
	   $\sum{p_T^l}$
           was calculated using all the leptons in the event.
           The quoted uncertainties consist of model uncertainties,
           MC statistical uncertainties and 
           systematic experimental uncertainties added in quadrature.
}
  \label{tab-comparison_sumpt100}
\end{table}
\begin{table}
  \begin{center}
     {\bf ZEUS} {\boldmath(${\mathcal L} = 480 \pbi$)}

    \begin{tabular}{|c|c|c|}
      \hline
        $\gamma \gamma$ sample 
        & $\sigma_{\rm DATA}^{96-07}$ (pb) 
        & $\sigma_{\rm SM}$ (pb)\\
      \hline
      $ee$ 
        & $0.64\pm 0.05^{+0.04}_{-0.03}$
        & $0.71\pm 0.02$ \\
      $\mu\mu$ 
        & $0.59\pm 0.07^{+0.07}_{-0.06}$
        & $0.69\pm 0.02$  \\
      Combined
        & $0.63\pm 0.04\pm 0.03$
        & $0.70\pm 0.02$ \\ 
      \hline
    \end{tabular}

  \end{center}
  \caption{Total cross section for $\gamma\gamma \rightarrow ee$ 
          and $\gamma\gamma \rightarrow \mu \mu$ samples,
           combined as explained in the text,
           compared with the predictions from the {\sc Grape} MC.
  }	
  \label{tab-xsec}
\end{table}
\begin{table}
  \begin{center}
     {\bf ZEUS} {\boldmath(${\mathcal L} = 480 \pbi$)}
    \begin{tabular}{|c|c|c|c|c|}
      \hline
    \multirow{2}{*}{Bin (\gev)}    & \multicolumn{3}{c|}{$\sigma_{\rm DATA}^{96-07}~(\fb/\gev)$} & $\sigma_{\rm SM}$  \\
      \cline{2-4}
                       & $e^+e^-$   & $\mu^+\mu^-$ & Combined & (\fb/\gev)  \\ \hline
 15 $<M_{12}<$ 25 & $23.0\pm 3.2~^{+1.6}_{-1.2}$  
                       & $32.7\pm 5.0~^{+3.7}_{-3.2}$  
                       & $25.4\pm 2.7 \pm 1.5$
                       & $30.4\pm1.0$ \\
 25 $<M_{12}<$ 40 & $19.1\pm 2.0~^{+1.6}_{-1.2}$  
                       & $12.6\pm 2.6~^{+1.4}_{-1.3}$  
                       & $16.3\pm 1.6 \pm 1.1$
                       & $19.8\pm0.7$ \\
 40 $<M_{12}<$ 60 & $3.8\pm 0.8~^{+0.3}_{-0.4}$  
                       & $2.5\pm 1.0~^{+0.3}_{-0.3}$  
                       & $3.3\pm 0.6 \pm 0.3$
                       & $3.0\pm0.1$ \\
 60 $< M_{12}<$100& $0.15\pm 0.11~^{+0.04}_{-0.03}$ 
                       & $0.21\pm 0.21~^{+0.03}_{-0.02}$  
                       & $0.17\pm 0.10 \pm 0.03$
                       & $0.26\pm 0.02$ \\
      \hline\hline
 10 $<p_T^{l_1}<$ 15 & $90.7\pm 8.4^{+6.1}_{-2.5}$  
                    & $94  \pm 12 ^{+11}_{-9}$ 
                    & $91.6\pm 6.9\pm 5.3$
                    &$103.2\pm 3.3$ \\
 15 $<p_T^{l_1}<$ 20 & $26.4\pm 4.3^{+2.2}_{-1.8}$  
                    & $10.7\pm 4.4^{+1.3}_{-1.1}$ 
                    & $18.1\pm 3.1 \pm 1.3$
                    & $23.7\pm 0.9$ \\
 20 $<p_T^{l_1}<$ 25 & $ 4.2\pm 1.7^{+0.8}_{-0.5}$
                    & $ 8.9\pm 4.0^{+1.0}_{-0.9}$
                    & $ 5.0\pm 1.6\pm 0.7$
                    & $ 7.3\pm 0.4$ \\
 25 $<p_T^{l_1}<$ 50 & $ 0.90\pm 0.37^{+0.08}_{-0.11}$
                    & $ 0.70\pm 0.50^{+0.08}_{-0.08}$
                    & $ 0.82\pm 0.29\pm 0.08$
                    & $ 0.88\pm 0.06$ \\
     \hline\hline
 15 $<\sum{p_T^l} <$ 25 & $36.7\pm 4.0^{+2.4}_{-1.3}$
                        & $38.4\pm 5.5^{+4.4}_{-3.9}$
                        & $37.2\pm 3.2\pm 2.2$
                        & $43.9\pm 1.4$ \\
 25 $<\sum{p_T^l} <$ 40 & $15.8\pm 1.9^{+1.3}_{-1.1}$
                        & $11.2\pm 2.6^{+1.3}_{-1.1}$
                        & $14.0\pm 1.5\pm 0.9$
                        & $14.6\pm 0.5$ \\
 40 $<\sum{p_T^l} <$ 60 & $ 1.24\pm 0.44^{+0.21}_{-0.17}$
                        & $ 1.32\pm 0.76^{+0.15}_{-0.14}$
                        & $ 1.26\pm 0.38\pm 0.16$
                        & $ 1.69\pm 0.11$ \\
60 $<\sum{p_T^l} <$ 100 & $ 0.092\pm 0.092^{+0.021}_{-0.024}$
                        & $ 0.18\pm 0.18^{+0.02}_{-0.02}$
                        & $ 0.11\pm 0.08\pm 0.02$
                        & $ 0.16\pm 0.02$ \\
     \hline
    \end{tabular}
  \end{center}
  \caption{Differential cross section as a function of  the invariant mass, $M_{12}$,
           the transverse momentum of the highest-$p_T$ lepton, $p_T^{l_1}$, and the
          scalar sum of the transverse momentum of the two leptons, $\sum{p_T^l}$,
          for di-lepton events in the kinematic region
          defined in the text, compared with the predictions from the {\sc Grape} Monte Carlo.
          The results are shown separately for the $ee$ and $\mu \mu$ samples, as well as for the combined sample.
}	
  \label{tab-xsec2}
\end{table}
\vfill\eject


\vfill\eject

\end{document}